\documentclass[a4paper,12pt]{article}
\usepackage{amsmath,amsfonts}
\begin{document}
\newcommand{\ds}{\displaystyle}
\newcommand{\be}{\begin{equation}}
\newcommand{\ee}{\end{equation}}
\newcommand{\ba}{\begin{array}}
\newcommand{\ea}{\end{array}}
\newcommand{\sn}{\mbox{sn}}
\newcommand{\cn}{\mbox{cn}}
\newcommand{\dn}{\mbox{dn}}
\newcommand{\bea}{\begin{eqnarray}}
\newcommand{\eea}{\end{eqnarray}}
\newcommand{\bi}{\begin{itemize}}
\newcommand{\ei}{\end{itemize}}
\newcommand{\x}{{\ensuremath{\times}}}
\newcommand{\bb}[1]{\makebox[16pt]{{\bf#1}}}
\newcommand{\stakel}{St\"ackel}
\newcommand{\ch}{\mathcal{H}}
\newcommand{\cl}{\mathcal{L}}
\newcommand{\cm}{\mathcal{M}}
\newcommand{\crr}{\mathcal{R}}
\newcommand{\cS}{\mathcal{S}}
\newcommand{\inv}{\frac{1}}
\newtheorem{theorem}{Theorem}
\newtheorem{definition}{Definition}
\newtheorem{lemma}{Lemma}
\newtheorem{comment}{Comment}
\newtheorem{corollary}{Corollary}
\newtheorem{example}{Example}
\newtheorem{examples}{Examples}

\title{Coupling constant metamorphosis and $N$th order symmetries in classical and quantum mechanics}
\author{E.\ G.\ Kalnins\\
{\sl Department of Mathematics and Statistics,}\\
{\sl University
of Waikato,}\\{\sl Hamilton, New Zealand.}\\
 W.\ Miller, Jr.
\\
{\sl School of Mathematics, University of Minnesota,}\\
{\sl Minneapolis, Minnesota,
55455, U.S.A.}\\
S. Post\\
{\sl School of Mathematics, University of Minnesota,}\\
{\sl Minneapolis, Minnesota,
55455, U.S.A.}
}
\date{\today}
\maketitle

\vspace{0.3cm}
%\tableofcontents
\begin{abstract}
We review the  fundamentals of coupling constant metamorphosis (CCM) and the St\"ackel transform, and 
apply them to map integrable and superintegrable systems of all orders into other
such  systems on different manifolds. In general, CCM does not preserve the order
of constants of the motion or even take polynomials in the momenta to
polynomials in the momenta.   We study specializations of these actions which do preserve
polynomials and also the structure of the symmetry algebras in both the classical and quantum cases.  We give several examples of non-constant curvature 3rd and 4th order superintegrable systems in 2 space dimensions obtained via CCM, with some details on the structure of the symmetry algebras preserved by the transform action.
\end{abstract}

\maketitle
\section{Introduction}  
There has been a recent rapid expansion in the number of known classical and quantum superintegrable systems of order 2, \cite{KKMP, KKMW2003}, and, particularly, of order 3 and higher, \cite{GW, Gravel,SCQS,Evans2008a,Evans2008b, TTW}. For many of these systems it has been demonstrated that the algebra generated by the fundamental higher order symmetries closes under the Poisson bracket in the classical case,  and under the commutator in the quantum case, to form a finite dimensional quadratic or cubic algebra. The representation theory of these algebras and their association with basic properties of the special functions of mathematical physics is of great current interest, \cite{Quesne, CDas, DASK2007,Quesne2, KMPost,IM2009, POST2009}. Indeed, the basic properties of Gaussian hypergeometric functions and their various limiting cases, as well as Lam\'e, Mathieu and Heun functions, and ellipsoidal harmonics all appear as associated with 2nd order superintegrable quantum systems via separation of variables. These functions as well as orthogonal polynomials of a discrete variable, including the general Wilson and Racah polynomials, are bound up with function space models of the irreducible representations of the quadratic algebras associated with 2nd order superintegrable quantum systems. The Painlev\'e transcendents (not associated with variable separability) appear in the study of 3rd order superintegrable systems. Some examples are known for conformally flat manifolds in $n$ dimensions, \cite{KWMP,HER1, BH}, but most results are known for 2 and 3 dimensional conformally flat spaces.

There is a disconnect, however, between what is known for 2nd order superintegrable systems and what is known for 3rd and higher order systems. For 2nd order superintegrable systems, classical and quantum, all such systems and all manifolds of dimension 2 on which they occur have been classified and the mechanism of the closure of the quadratic algebra is well understood, \cite{KKM20041,KKM20042,KKM20061, DASK2005,KKMP2009}. For conformally flat manifolds in 3 dimensions  great advances have been made although the classification and structure analysis is not yet complete, \cite{KKM20051, KKM20052,KKM20061,KKM2007,KKM2007b}. A major tool for obtaining these 2nd order results has been the St\"ackel transform \cite{BKM}, a variant of coupling constant metamorphosis \cite{HGDR}, which enables a 1-1 invertible transformation between a 2nd order superintegrable system on one manifold and a superintegrable system on another manifold that preserves the symmetry algebra structure. This has given us an elegant method for classification of all 2D  superintegrable systems through the important fact that every such system can be shown to be the St\"ackel transform of a system on a constant curvature space, \cite{KKMW2003, KKM20042,KKMP2009}. Also it gives important insight into the structure of Koenigs' remarkable potential-free results \cite{Koenigs}. Similar results are known for 3D systems but the classification is not yet complete, 
\cite{KKM20052, KKM2007, KKM2007b}.

For 3rd and higher order superintegrable systems, however, there is no structure and classification theory. Only examples are known, and these are very difficult to obtain. The symmetry algebras can be computed for each example  but the mechanism for their closure and structure is not understood. Virtually all known examples are in 2D or 3D Euclidean space. The present paper is a first attempt at refining a tool (CCM/St\"ackel transform), that has  proved so successful in the classification and structure theory for 2nd order systems, so that it applies to higher order superintegrable systems. There are two basic issues  here. The first is that CCM in general doesn't preserve the structure of the symmetry algebras. We have to determine a suitable restriction that does preserve the structure. Secondly, CCM is a classical phenomenon; its extension to the quantum case is not automatic and requires special care. In this paper most of our classical results will be stated for $n$ dimensional systems whereas, for simplicity,  the quantum results will be limited mostly to 2 dimensions. 

In future papers we will extend the operator CCM to 3 and higher dimensions and employ this tool to attack the structure and classification theory for 3rd and higher order superintegrable systems in all dimensions. An immediate result of the present paper is the explicit display of a large number of higher order superintegrable systems on manifolds not of constant curvature, the existence of which seems not to be widely recognized. We also provide new examples of explicit structure computations for the quadratic algebras of some 3rd and 4th order superintegrable systems on 2D Euclidean space that map to isomorphic systems on nonconstant curvature spaces.

Before proceeding to our results we give some basic definitions that we employ throughout the paper.
 A classical
superintegrable system on an $n$-dimensional real or complex Riemannian or pseudo-Riemannian manifold is defined by its associated Hamiltonian function
 ${\cal H} =\sum_{ij}g^{ij}p_ip_j+V({\bf x})$ on
the phase space of this manifold. Here $g^{ij}({\bf x})$ is the contravariant metric tensor in local coordinates $\bf x$ and $V({\bf x})$ is a prescribed function that may depend on some parameters.  The system  is {\bf superintegrable} if it 
admits $2n-1$ functionally independent  generalized
symmetries (or constants of the motion) ${\cal S}_k,\quad k=1,\cdots,2n-1$ with ${\cal S}_1={\cal H}$ where the ${\cal S}_k$  are polynomials in the momenta $p_j$.  That is, $\{{\cal H},{\cal S}_k\}=0$ where 
$ \{f,g\}=\sum_{j=1}^n(\partial_{x_j}f\partial_{p_j}g-\partial_{p_j}f\partial_{x_j}g)$
is the Poisson bracket for functions $f({\bf x},{\bf p}),g({\bf
  x},{\bf p})$ on phase space.  It is easy to see that $2n-1$ is the maximum possible 
number of functionally independent symmetries and, locally, 
such (in general nonpolynomial) symmetries always exist.   Most authors, but not us, also demand that the system is {\bf integrable}, i.e., there is a subset of $n$ functionally independent polynomial symmetries, say ${\cal S}_1,\cdots, {\cal S}_n$, such that  $\{{\cal S}_j,{\cal S}_\ell\}=0$, $1\le s,\ell\le n$. If the maximum order of the polynomials corresponding to the generating symmetries is $N$, we say that the system is $N$th order superintegrable.

Superintegrable systems can lay claim to be the most symmetric
Hamiltonian  systems  though
many such systems admit no group symmetry; the symmetry is ``hidden''.  Generically, every
geometrical trajectory  in phase space (but not the time dependence of the trajectory  ${\bf p}(t),{\bf x}(t)$)
of the Hamilton equations of motion for the system, is obtained as the
common intersection of the (constants of the motion) hypersurfaces 
${\cal S}_k({\bf p},{\bf x})=c_k,\quad k=0,\cdots,2n-2$.
The orbits can be found without solving the equations of
motion. Since every known superintegrable system is also integrable,
this is better than integrability. A case can be made that the  2nd order superintegrability of the
Kepler-Coulomb two-body 
problem, forcing the existence of conic sections as trajectories,  is the reason that Kepler was able to determine the planetary
elliptical orbits before the invention of calculus.

There is an analogous definition of superintegrability for quantum
 systems with Schr\"odinger operator 
\[ H=\Delta +V({\bf x}),\quad
\Delta=\frac{1}{\sqrt{g}}\sum_{ij}\partial_{x_i}(\sqrt{g}g^{ij})\partial_{x_j},
\]
  the Laplace-Beltrami operator  plus a
potential function. Here it is required that there are $2n-1$
  functionally independent  differential operators, $S_1=H,S_2,\cdots,S_{2n-1}$ such that 
and $[H,S_k]\equiv HS_k-S_kH=0$.  Often  there is a 1-1 relationship between  classical and quantum superintegrable systems associated with a potential and then functional independence refers to the classical system. In those cases where there is no classical analog, however, there is no agreed upon definition of quantum functional independence.
A basic motivation for studying these  systems is that they can be  solved explicitly, often in multiple ways. Typically their symmetry algebras close to form quadratic, cubic, or similar algebras whose representation theory yields spectral information about the quantum system. 

In the following sections we review the basic definition and properties of coupling constant metamorphosis (CCM) and the closely related St\"ackel transform for classical systems. These concepts apply to any Hamiltonian system with potential, not just superintegrable systems. Then we define specializations of these general concepts that preserve the order of symmetries and also define symmetry algebra isomorphisms. It is these specializations that are needed for the study of superintegrable systems.  Then, and most importantly, we find quantum analogs of these classical transforms. At each stage we provide examples, several of them new. 

\section{Coupling constant metamorphosis}
The basic tool that we will employ follows from ``coupling constant
metamorphosis" (CCM), a general fact about Hamiltonian systems, pointed out in
\cite{HGDR}.
Let ${\cal H}({\bf x},{\bf p})+\alpha U({\bf x})$ define a Hamiltonian system
in 2$n$ dimensional phase space, with canonical coordinates $x_j,p_j$. Thus the
Hamilton-Jacobi equation would take the form ${\cal H}({\bf x},{\bf p})+\alpha
U({\bf x})=E$. Assume that for every value of the parameter $\alpha$ the system
admits a constant of the motion ${\cal K}(\alpha)$, analytic in $\alpha$.
\begin{theorem}\label{theorem0} Coupling constant metamorphosis. The
Hamiltonian ${\cal H}'=({\cal H}-E)/U$ admits the constant of the motion
${\cal K}'={\cal K}(-{\cal H}')$, where now $E$ is a parameter. \end{theorem}

\medskip\noindent PROOF: Note that if $F,G$ are functions on phase space of the
form $G({\bf x}, {\bf p})$, $F=F(a)=F(a,{\bf x},{\bf p})$ where $a=\alpha({\bf
x},{\bf p})$ then
$$\{F,G\}= \{F(a),G\}|_{a=\alpha({\bf x},{\bf p})}+\partial_a F(a)
|_{a=\alpha({\bf x},{\bf p})}\{\alpha,G\}.$$
By assumption, $\{{\cal K}(\alpha),{\cal H}\}=-\alpha\{{\cal K}(\alpha),U\}$
for any value of the parameter $\alpha$. Thus
$$\{{\cal K}(\alpha),{\cal H}'\}=\frac{\{U,{\cal K}(\alpha)\}}{U}({\cal
H}'+\alpha).$$
Now
$$\{{\cal K}({\cal H}'),{\cal H}'\}=\left[\partial_{\alpha} {\cal
K}(\alpha)\{{\cal H}',{\cal H}'\}+\frac{\{U,{\cal K}(\alpha)\}}{U}({\cal
H}'+\alpha)\right]_{\alpha=-{\cal H}'}=0.$$
Q.E.D

\begin{corollary} Let ${\cal K}_1(\alpha), {\cal K}_2(\alpha)$ be constants of
the motion for the system  ${\cal H}({\bf x},{\bf p})+\alpha U({\bf x})$. Then 
$\{{\cal K}_1, {\cal K}_2\}(\alpha)\equiv \{{\cal K}_1(\alpha), {\cal
K}_2(\alpha)\}$ is also a constant of the motion and
$$ \{{\cal K}_1(-{\cal H}'), {\cal K}_2(-{\cal H}')\}=\{{\cal K}_1, {\cal
K}_2\}(-{\cal H}'). $$
\end{corollary}

Clearly CCM takes integrable systems to integrable systems and superintegrable
systems to superintegrable systems. We are concerned with the case where
\be\label{hamiltonian} {\cal H}=\sum_{i,j=1}^ng^{ij}p_ip_j+V({\bf x})+\alpha U({\bf x})\equiv {\cal
H}_0+V+\alpha U\ee
is a classical Hamiltonian system on an $n$-dimensional pseudo-Riemannian
manifold and are interested only in those constants of the motion $\cal K$ that
are polynomial in the momenta. As we shall see, in the  case of 2nd order
constants of the motion  there is special structure. The 2nd order constants
of the motion are typically at most linear in $\alpha$, so they transform to 2nd
order symmetries again. In this case CCM agrees with the St\"ackel transform
that we shall take up in the next section. However, in general the order of
constants of the motion is not preserved by coupling constant metamorphosis.
\begin{example}\label{example1} The system
$$ {\cal H}= p_1^2+p_2^2 + b_1\sqrt{x_1}+b_2x_2$$
admits the 2nd order constant of the motion ${\cal K}^{(2)}=p_2^2+b_2x_2$ and
the 3rd order constant of the motion ${\cal K}^{(3)}=p_1^3+\frac32
b_1\sqrt{x_1}p_1-\frac{3b_1^2}{4b_2}p_2$, (\cite{Gravel} and references
contained therein).
If we choose  $\alpha U=\alpha\sqrt{x_1}$ then the transform of ${\cal K}^{(3)}$
will be 5th order. If we choose $\alpha U=\alpha x_2$ then the transform of
${\cal K}^{(3)}$ will be rational, but  nonpolynomial. Thus to obtain useful
structure results from this general transform, and to obtain results that have
the possibility of carrying over to the quantum case, we need to restrict the
generality of the transform action.
\end{example}

\section{The Jacobi transform}
Here we study a specialization of coupling constant metamorphosis  to the
 case where $V=0$.  The special version of the transform we study  takes
$N$th order constants of the motion for Hamiltonian systems to $N$th order
constants of the motion.
An $N$th order constant of the motion ${\cal K}({\bf x},{\bf p})$ for the
system
\be\label{hamiltonian1} {\cal H}=\sum_{i,j=1}^ng^{ij}p_ip_j+U({\bf x})={\cal
H}_0+U\ee
 is a function on the phase space such that
$\{{\cal K},{\cal H}\}=0$ where
$$ {\cal K}={\cal K}_N+{\cal K}_{N-2}+{\cal K}_{N-4}+\cdots+{\cal K}_0,\quad  
n\ {\rm even},$$
$$ {\cal K}={\cal K}_N+{\cal K}_{N-2}+{\cal K}_{N-4}+\cdots+{\cal K}_1,\quad  
n\ {\rm odd}.$$
Here, ${\cal K}_N\ne 0$ and ${\cal K}_j$ is homogeneous in $\bf p$ of order
$j$.
This implies the conditions
\be\label{killingtensor} \{{\cal K}_N,{\cal H}_0\}=0,\ee
\be \label{cond1}\{{\cal K}_{N-2k},U\}+\{{\cal K}_{N-2k-2},{\cal H}_0\}=0,
\quad k=0,1,\cdots,[N/2]-1,\ee
and, for $N$ odd,
\be\label{cond2}  \{{\cal K}_1,U\}=0.\ee

The case $N=1$ is very special. Then ${\cal K}={\cal K}_1$ and the conditions
are
$$\{ {\cal K}, {\cal H}_0\}=0,\quad \{{\cal K},U\}=0,$$
so $\cal K$ is a Killing vector and $U$ is invariant under the local group
action generated by the Killing vector.

For $N=2$, ${\cal K}={\cal K}_2+{\cal K}_0$ and the conditions are
\be\label{N=2}\{ {\cal K}_2, {\cal H}_0\}=0,\quad \{{\cal K}_2,U\}+\{{\cal
K}_0,{\cal H}_0\}=0,\ee
so ${\cal K}_2$ is a 2nd order Killing tensor and $U$ satisfies (linear)
Bertrand-Darboux integrability conditions.

For $N=3$, ${\cal K}={\cal K}_3+{\cal K}_1$ and the conditions are
$$\{ {\cal K}_3, {\cal H}_0\}=0,\quad \{{\cal K}_3,U\}+\{{\cal K}_1,{\cal
H}_0\}=0,\quad \{{\cal K}_1,U\}=0 $$
so ${\cal K}_3$ is a 3rd order Killing tensor. The integrability conditions for
the last 2 equations lead to nonlinear PDEs for $U$.

\begin{theorem}\label{theorem1} Suppose  the system (\ref{hamiltonian1}) admits
an $N$th order constant of the motion ${\cal K}$ where $N\ge 1$.  Then
$${\hat{\cal K}}=\sum_{j=0}^{[N/2]}\left(-\frac{{\cal H}_0-E}{U}\right)^j{\cal
K}_{N-2j}$$
 is an $N$th order constant of the motion for the  system $({\cal
H}_0-E)/U$.\end{theorem}

\medskip\noindent PROOF:  It follows from the  general conditions
(\ref{killingtensor}), (\ref{cond1}), (\ref{cond2}). That
$${\cal K}(\alpha)= \sum_{j=0}^{[N/2]} \alpha^j {\cal K}_{N-2j}$$
is a constant of the motion for the system
$ {\cal H}(\alpha)={\cal H}_0+\alpha U$.
Then from Theorem \ref{theorem0}, we have that  ${\cal K} (-\frac{{\cal
H}_0-E}{U})$ is an $N$th order constant of the motion for the system $({\cal
H}_0-E)/U$.
Q.E.D.

Note that if we set $E=0$ then  $\hat{\cal K}$ becomes an $N$th order Killing
tensor for the free system ${\cal H}_0/U$.
\begin{corollary}\label{cor2} Suppose the system ${\cal H}_0+U$ is $N$th  order
superintegrable. Then the free system ${\cal H}_0/U$ is also $N$th order
superintegrable.\end{corollary}

We will call  ${\hat {\cal K}}$ a {\it Jacobi transform} of ${\cal K}$, in
recognition of its close relationship to the Jacobi metric, \cite{LW} page 172,
and to distinguish it from the St\"ackel transform and more general coupling
constant metamorphosis. Note that the Jacobi transform for general parameter
$E$ is invertible.

Corollary \ref{cor2} tells us that each of the 3rd order superintegrable
systems found by Gravel in 2D Euclidean space,\ \cite{Gravel},  yields  superintegrable systems
on conformally flat manifolds, usually not of constant curvature.

\begin{corollary}\label{cor3} The Jacobi transform satisfies the properties
 $$ 
\widehat{\{{\cal K},{\cal L}\}}=\{\hat{\cal K},\hat{\cal L}\},\ \widehat{{\cal
K}{\cal L}}=\hat{\cal K}\hat{\cal L},$$
and, if ${\cal K},\ {\cal L}$ are of the same order, $\widehat{a{\cal K}+b{\cal L}}=a\hat{\cal K}+b\hat{\cal L}.$
Thus it defines a homomorphism from the graded symmetry algebra of the system ${\cal
H}_0+U$ to the graded symmetry algebra of the  system $({\cal H}_0-E)/U$.
\end{corollary}
\begin{example}\label{example2}
Consider the  system of Example \ref{example1}:
$ {\cal H}= p_1^2+p_2^2 + b_1\sqrt{x_1}+b_2x_2$, and let $U=
b_1\sqrt{x_1}+b_2x_2+b_3$ for some fixed $b_1,b_2,b_3$ with $b_1b_2\ne 0$. The new Hamiltonian is
$$\hat{\cal H}= \frac{p_1^2+p_2^2-E}{b_1\sqrt{x_1}+b_2x_2+b_3}.$$
and the Jacobi transforms of ${\cal K}^{(2)},{\cal K}^{(3)}$ are
$$\hat{{\cal
K}}^{(2)}=p_2^2-b_2x_2\left(\frac{p_1^2+p_2^2-E}{b_1\sqrt{x_1}+b_2x_2+b_3}\right),$$
$$ \hat{{\cal K}}^{(3)}=p_1^3-(\frac32
b_1\sqrt{x_1}p_1-\frac{3b_1^2}{4b_2}p_2)\left(\frac{p_1^2+p_2^2-E}{b_1\sqrt{x_1}+b_2x_2+b_3}\right).$$
\end{example}

\section{The St\"ackel transform}
Using the same notation as in the previous section, and a particular nonzero
potential $U=V({\bf x},{\bf b}_0)$
 we define the St\"ackel transform  for a system ${\cal H}={\cal H}_0+V({\bf
x},{\bf b})$, \cite{BKM}.   The transform of ${\cal K}={\cal  K}_1$ is
$\tilde{\cal K}={\cal K}_1$ The transform of ${\cal K}={\cal K}_2+{\cal K}_0$
is $\tilde {\cal K}={\cal K}-\frac{{\cal K}_0^U}{U}{\cal H}$. (Here ${\cal
K}_j$ is a homogeneous polynomial in $\bf p$ of order $2j$, and ${\cal K}_j^U$
is the restriction of ${\cal K}_j$ to the potential $V=U$.) The transform maps
1st and 2nd order constants of the motion for $\cal H$ to constants of the
motion for the system ${\cal H}/U$. Thus the system $\cal H$ is 2nd
order superintegrable iff the system ${\cal H}/U$ is 2nd order
superintegrable. For completeness we review briefly the direct  proofs of the
basic theoretic facts.
 \begin{theorem} Let $\cal K$ be a 2nd order constant of the motion for the
system $\cal H$ and $U$ be a particular instance of the potential $V$. Then
$\tilde{\cal K}$ is a 2nd order constant of the motion for the system
${\cal H}/U$. \end{theorem}

 \medskip\noindent PROOF:  $$\{\tilde{\cal K},\frac{\cal H}{U}\}=\{{\cal
K}-\frac{{\cal K}_0^U}{U}{\cal H},\frac{\cal H}{U}\}=-\frac{\cal
H}{U^2}\left(\{{\cal K}_2,U\}+\{{\cal K}_0^U,{\cal H}_0\}\right)=0$$
Q.E.D.

\begin{corollary}\label{corollary2} Let ${\cal K}, {\cal L}$ be 2nd order
constants of the motion for the system $\cal H$ and let $\tilde{\cal K},\
\tilde{\cal L}$ be their respective St\"ackel transforms determined by the 
potential $U$. If $\{{\cal K},{\cal L}\}=0$ then $\{\tilde{\cal K},\tilde{\cal
L}\}=0$.\end{corollary}

\medskip\noindent PROOF:  Suppose $\{{\cal K},{\cal L}\}=0$. We have
$$\{{\cal K},{\cal L}\}=\{{\cal K}_2,{\cal L}_2\} +\left(\{{\cal K}_2,{\cal
L}_0\}+\{{\cal K}_0,{\cal L}_2\}\right)=0,$$
where the first term on the right is of order 3 and the second term is of order
1. Thus
$$\{{\cal K}_2,{\cal L}_2\} =\{{\cal K}_2,{\cal L}_0\}+\{{\cal K}_0,{\cal
L}_2\}=0.$$
Then, a straightforward computation yields
$$\{\tilde{\cal K},\tilde{\cal L}\}=\{{\cal K},{\cal L}\}-\frac{\cal
H}{U}\left(\{{\cal K}_2,{\cal L}_0^U\}+\{{\cal K}_0^U,{\cal L}_2\}\right)=0.$$
Q.E.D.

\begin{corollary}
\label{corollary3} Let $\{{\cal K},{\cal L}\}=0$ be as in Corollary
\ref{corollary2} and assume that one instance of the potential $V$ is the
constant $1$, i.e., $V({\bf b}_1)=1$. Then if $\{\tilde{\cal K},\tilde{\cal
L}\}=0$ we must have $\{{\cal K},{\cal L}\}=0$.\end{corollary}

\medskip\noindent PROOF: Suppose $\{\tilde{\cal K},\tilde{\cal L}\}=0$. Then
the order 3 and order 1 terms on the left hand side of this expression must
vanish separately:
\be\label{3rdorder}\{{\cal K}_2,{\cal L}_2\}-\frac{{\cal H}_0}{U}\left(\{{\cal
K}_2,{\cal L}_0^U\}+\{{\cal K}_0^U,{\cal L}_2\}\right)=0,\ee
\be\label{1storder}\{{\cal K}_2,{\cal L}_0\}+\{{\cal K}_0,{\cal
L}_1\}-\frac{V}{U}\left(\{{\cal K}_2,{\cal L}_0^U\}+\{{\cal K}_0^U,{\cal
L}_2\}\right)=0,\ee
Identity (\ref{3rdorder}) says that
\be\label{commident1}\{{\cal K}_2,{\cal L}_2\}={\cal H}_0{\cal X}\ee
 where ${\cal X}=\frac{1}{U}\left(\{{\cal K}_2,{\cal L}_0^U\}+\{{\cal
K}_0^U,{\cal L}_2\}\right)$. Since ${\cal K}_2,{\cal L}_2$ are 2nd order
Killing tensors of ${\cal H}_0$, it follows easily from the Jacobi relation for
the Poisson bracket that $\cal X$ is a Killing vector. From identity
(\ref{1storder}) we obtain the result
\be\label{commident2} \{{\cal K},{\cal L}\}+{\cal H}{\cal X}=\{\tilde{\cal
K},\tilde{\cal L}\}=0.\ee
 Taking the Poisson bracket of the left hand side of this last identity with
$\cal H$ we see that $\cal X$ is a first order constant of the motion for
system $\cal H$.
From (\ref{1storder}) we have
$${\cal X}=\frac{1}{U}\left(\{{\cal K}_2,{\cal L}_0^U\}+\{{\cal K}_0^U,{\cal
L}_2\}\right)=\frac{1}{V}\left(\{{\cal K}_2,{\cal L}_0^V\}+\{{\cal K}_0^V,{\cal
L}_2\}\right)$$
for any nonzero choice of potential $V$. Choosing $V=1$ we find
\be\label{X1} {\cal X}=\{{\cal K}_2,{\cal L}_0^1\}+\{{\cal K}_0^1,{\cal
L}_2\}.\ee
From relation ({\ref{N=2}}) with $V=1$ we have $\{{\cal K}_2,1\}+\{{\cal
K}_0^1,{\cal H}_0\}=0$ so $\{{\cal K}_0^1,{\cal H}_0\}=0$. Since the metric is
nondegenerate, this implies that ${\cal K}_0^1=c_1$, a constant. Similarly,
${\cal L}_0^1=c_2$ is constant. Thus (\ref{X1}) implies ${\cal X}=0$, which
together with (\ref{commident2}) implies $\{{\cal K},{\cal L}\}=0$. Q.E.D.

An alternate way of proving Corollary \ref{corollary3} is to demonstrate that
there is an ``inverse"  St\"ackel transform that takes the system ${\cal
H}/U$ to $\cal H$ via the special potential ${1}/{U}$. The outcome of
applying the initial transform to a 2nd order constant of the motion $\cal K$
of $\cal H$ and then transforming back is ${\cal K}-{\cal K}_0^1{\cal H}$,
where ${\cal K}_0^1$ is a constant. If each 2nd order symmetry $\cal K$ is
normalized by the requirement ${\cal K}_0^1=0$, (by adding a suitable constant)
then this action is the identity operator.

These results show that the St\"ackel transform takes 2nd order
superintegrable systems to 2nd order superintegrable systems, preserves
variable separability, and is invertible. As stated in this generality for
second order symmetries, the St\"ackel transform is not a special case of
coupling constant metamorphosis, although the two transforms are closely
related. However in the situation where the potential functions $V({\bf x},{\bf
b})$ form a finite dimensional vector space,   which is usual in the study of
2nd order superintegrability, then the
transforms coincide. In this case, by redefining parameters if necessary, we
can assume $V$ is linear in  $\bf b$.

Now we will investigate extensions of the St\"ackel transform to higher order
constants of the motion, under the assumption that $V({\bf x},{\bf b})$  is
linear in ${\bf b}=(b_0,b_1,\cdots b_M)$, $U$ is of the form $U({\bf x}) =
V({\bf x},{\bf b}^0)$ and the potentials $V({\bf x}, {\bf b})$ span a space of
dimension $M+1$.
In particular,
\be \label{Spot}V({\bf x},{\bf b})=b_0+\sum_{i=1}^M U^{(i)}({\bf x})b_i\ee
where the set of functions $\{1,U^{(1)}({\bf x}),\cdots,U^{(M)}({\bf x})\}$ is
linearly independent.
In the study of 2nd order superintegrability, typically the 2nd order
constants of the motion are linear in the $\bf b$ and the algebra generated by
these symmetries via products and commutators has the property that a constant
of the motion of order $N$ depends polynomially on the parameters with order
$\le [N/2]$.
Thus we  consider only those higher order constants of the motion of order $N$
of the form
\be\label{Ssymm}{\cal K}=\sum_{j=0}^{[N/2]} {\cal K}_{N-2j}({\bf p}, {\bf
b})\ee
where ${\cal K}_{N-2j}(a{\bf p},{\bf b})=a^{N-2j}{\cal K}_{N-2j}({\bf p},{\bf
b})$ and ${\cal K}_{N-2j}({\bf p},a{\bf b})=a^{j}{\cal K}_{N-2j}({\bf p},{\bf
b})$ for any parameter $a$.
Let ${\cal K}({\bf b})$ be such an $N$th order constant of the motion.
Then
\be\label{Ssym1}{\cal K}(\alpha)\equiv {\cal K}({\bf p},{\bf b} +\alpha {\bf
b}^{(0)})\ee
 is an $N$th order constant of the motion for the system with Hamiltonian
${\cal H}_0+V({\bf x},{\bf b})+\alpha U({\bf x})$.
 Applying Theorem \ref{theorem0} we have
 \begin{theorem} Let ${\cal K}$ be an $N$th order constant of the motion for
the system ${\cal H}_0+V({\bf x},{\bf b})$ where $V$ is of the form
(\ref{Spot}) and $\cal K$ is of the form (\ref{Ssymm}). Let ${\cal K}(\alpha)$ be defined by 
(\ref{Ssym1}). Then
$$\tilde{\cal K}={\cal K}\left(-\frac{{\cal H}_0 +V({\bf x},{\bf b})}{U({\bf
x})}\right)=\sum_{j=0}^{[N/2]}\tilde{{\cal K}}_{N-2j}({\bf p},{\bf b}) $$
is an $N$th order constant of the motion for the system $({\cal H}_0 +V({\bf
x},{\bf b}))/{U({\bf x})}$, where
\be\label{Ssym2}\tilde{{\cal K}}_{N-2j}(a{\bf p},{\bf b})=a^{N-2j}\tilde{{\cal
K}}_{N-2j}({\bf p},{\bf b}),\quad \tilde{{\cal K}}_{N-2j}({\bf p},a{\bf
b})=a^{j}\tilde{{\cal K}}_{N-2j}({\bf p},{\bf b})\ee
 for any parameter $a$.
\end{theorem}

\begin{example}\label{example3}
This example of a 4th order superintegrable system is taken from \cite{TTW} and corresponds  to the choice $k=2$ for the potential
 $ V=Ar^2+B/r^2\cos^2 (kt)+C/r^2\sin^2 (kt)$ for suitable $A,B,C$, as  written in polar coordinates. The structure relations and transform are new.
Let
$$ {\cal H}=p^2_1+p^2_2+a(x_1^2+x_2^2)+b \frac{(x_1^2+x_2^2)}{ (x_1^2-x_2^2)^2} +c 
\frac{(x_1^2+x_2^2)}{ x_1^2x_2^2}$$
There are  two basic constants of the motion, one of 2nd order,
$${\cal K}_2=(x_1p_2-x_2p_1)^2+4b \frac{x_1^2x_2^2}{(x_1^2-x_2^2)^2} + c \frac{(x_1^4+x_2^4)}{ x_1^2x_2^2}$$
and one of 4th order,
$${\cal K}_4=(p^2_1-p^2_2)^2+[2ax_1^2+ 2b \frac{(x_1^2+x_2^2)}{(x_1^2-x_2^2)^2} -2c 
\frac{(x_1^2-x_2^2)}{ x_1^2x_2^2} ]p^2_1$$
$$+[-4ax_1x_2+ 8b \frac{x_1x_2}{ (x_1^2-x_2^2)^2} ]p_1p_2+[2ax_2^2+ 2b 
\frac{(x_1^2+x_2^2)}{ (x_1^2-x_2^2)^2} +2c \frac {(x_1^2-x_2^2)}{ x_1^2x_2^2} ]p^2_2$$
$$+a^2(x_1^2-x_2^2)^2+ \frac{b^2}{ (x_1^2-x_2^2)^2} + c^2 \frac{(x_1^2-x_2^2)^2}{ x_1^4x_2^4}+8ab 
\frac{x_1^2x_2^2}{ (x_1^2-x_2^2)} +2 \frac{bc}{ x_1^2x_2^2}$$
These constants of the motion generate a closed Poisson algebra. Let
 ${\cal R}=\{{\cal K}_2,{\cal K}_4\}$.The  relations are 
$$\{{\cal K}_2,{\cal R}\}=32({\cal H}^2-2{\cal K}_4){\cal K}_2-64(b+2c){\cal K}_4+64(b-c){\cal H}^2-128ab{\cal K}_1-128ab(b+2c),$$
$$\{{\cal K}_4,{\cal R}\}=32{\cal K}_4({\cal K}_4-{\cal H}^2)+128a{\cal K}_2{\cal H}^2-384a^2{\cal K}^2_2+128ab{\cal K}_4-64(b+4c)a{\cal H}^2$$
$$+256a^2
(2c-b){\cal K}_2 + 128a^2(b^2+40c^2+20bc)$$
There is a Casimir constraint 
$${\cal R}^2=64{\cal K}_2{\cal K}_4({\cal H}^2-{\cal K}_4)-64b{\cal H}^4+128(b-c){\cal K}_4{\cal H}^2-64(b+2c){\cal K}^2_4-128a{\cal K}^2_2{\cal H}^2+$$
$$256a^2{\cal K}^3_2-256ab{\cal K}_2{\cal K}_4+128a(b+4c){\cal H}^2{\cal K}_2+256a^2(b-c){\cal K}^2_2-256ab(b+2c){\cal K}_4$$
$$+256a(7bc+b^2-2c^2){\cal H}^2-256a^2(b^2+4c^2+20bc){\cal K}_2-256a^2(2c+b)(b^2+16bc-4c^2).$$
Then the St\"ackel transformed system
$${\tilde {\cal H}}=\frac{p^2_1+p^2_2+a(x_1^2+x_2^2)+b \frac{(x_1^2+x_2^2)}{ (x_1^2-x_2^2)^2} +c 
\frac{(x_1^2+x_2^2)}{ x_1^2x_2^2}+d}{(x_1^2+x_2^2)+B \frac{(x_1^2+x_2^2)}{ (x_1^2-x_2^2)^2} +C 
\frac{(x_1^2+x_2^2)}{ x_1^2x_2^2}+D }$$
is also superintegrable with 4th and 2nd order generating constants of the motion.
Note that for $B=C=0$, $D=4$, the transformed system is defined on a Darboux space of type 3, whereas if $B=D=0$, $C=1$ the transformed system is defined on a Darboux space of type 2, \cite{KKMW2003}.
\end{example}

\section{2D  quantum symmetries}
Here we begin the study of quantum symmetries. The quantization is  much
simpler in the 2D  case than
for dimensions greater than 2, and for 1st and 2nd order symmetries, so we
begin with these special cases to gain insight.  Here the metric, expressed in
Cartesian-like coordinates, is
$ds^2=\lambda({\bf x})(dx_1^2+dx_2^2)$, and the  Hamiltonian system $
{\cal H}=(p_1^2+p_2^2)/\lambda({\bf x})+V({\bf x})$ is replaced by
the Hamiltonian (Schr\"odinger) operator with potential
\be \label{2DHamq}
H=\frac{1}{\lambda(\bf x)}(\partial_{11}+\partial_{22})+V({\bf x}).
\ee
\subsection{2nd order operator symmetries}
 A 2nd order symmetry of the
Hamiltonian system\
${\cal K}=\sum_{k,j=1}^2a^{kj}({\bf x})p_kp_j+W({\bf x})$, with
$a^{kj}=a^{jk}$,
corresponds to the  operator
$$ K=\frac{1}{\lambda({\bf x})}\sum_{k,j=1}^2\partial_k\left(\lambda({\bf
x})a^{kj}({\bf x})\partial_j\right)+W({\bf x}), \quad
a^{kj}=a^{jk}.
$$
These operators are formally self-adjoint with respect to the bilinear
product
$$ <f,g>_\lambda=\int f({\bf x})g({\bf x})\lambda({\bf x}) dx_1 dx_2
$$
on the manifold, i.e.,
$$ <f,Hg>_\lambda=<Hf,g>_\lambda,\quad <f,Kg>_\lambda=<Kf,g>_\lambda$$
for all local $C^\infty$ functions $f,g$ with compact support on the
manifold, where the domain of integration is ${\cal C}^2$ or ${\cal R}^2$. If the functions defining a differential operator are singular on a 1-dimensional or 0-dimensional set, we we restrict the support of $f,g$ to be bounded away from this set. We define the formal adjoint $T^*$ of a linear operator $T$ on the space  $C^\infty_0$ by
\be\label{innerproduct} <T^*f,g>_\lambda=<f,Tg>_\lambda\ee
for all $f,g\in C^\infty_0$. The operators $H,K$ are formally self-adjoint: $H^*=H,K^*=K$.

If the  Schr\"odinger equation admits a multiplicative separable solution in particular coordinates $x_1,x_2$ then the Schr\"odinger operator can  be written as
\be\label{sepham} H=\frac{1}{X^{(1)}(x_1)+X^{(2)}(x_2)}\left(\partial_{11}+
\partial_{22}+V^{(1)}(x_1)+V^{(2)}(x_2)\right)\ee
where the 2nd order symmetry responsible for the separation is
\be\label{sepsym} K=\frac{1}{X^{(1)}(x_1)+X^{(2)}(x_2)}\left(X^{(2)}(x_2)\partial_{11}-X^{(1)}(x_1)\partial_{22}\right.\ee
$$+\left. X^{(2)}(x_2)V^{(1)}(x_1)-X^{(1)}(x_1)V^{(2)}(x_2)\right).$$
Thus the metric is $\lambda({\bf x})=X^{(1)}(x_1)+X^{(2)}(x_2)$ and the potential is $V({\bf x})=(V^{(1)}x_1)+V^{(2)}x_2))/(X^{(1)}(x_1)+X^{(2)}(x_2))$.

A 1st order symmetry of the Hamiltonian system ${\cal
  L}=\sum_{k=1}^2a^k({\bf x})p_k$ corresponds to the operator
$$L=\sum_{k=1}^2\left(a^k({\bf x})\partial_k+\frac{\partial_k(\lambda({\bf
x})a^k({\bf x}))}{2\lambda({\bf x})}\right).$$
It is easy to show that $L$ is formally skew-adjoint, i.e.,  $L^*=-L$.

The following results that relate the operator commutator
$[A,B]=AB-BA$ and the
Poisson bracket are straightforward to verify.
\begin{lemma}\label{2ndorderop} $$\{ {\cal H},{\cal K}\}=0\ \iff \ [H,K]=0.$$
\end{lemma}
This result is not generally true for higher dimensional manifolds.
\begin{lemma} $$\{ {\cal H},{\cal L}\}=0\ \iff \ [H,L]=0.$$
\end{lemma}
The classical St\"ackel transform for these systems can easily be extended to
the operator case. Suppose $V$ is a parametrized potential and let $U$ be a
special instance of that potential. Let $K=\frac{1}{\lambda}\sum
\partial_i(\lambda a^{ij}\partial_j)+W=K_2+K_0$, where $K_0=W$,  be a
2nd order formally self-adjoint symmetry operator of $H$
and
$K^U_0$  be the restriction of $K_0$ to $V=U$. Then
$${\tilde K}=K-K_0^U U^{-1}H$$
is the corresponding formally self-adjoint symmetry operator of
${\tilde H}=U^{-1}H$, with respect to the metric $d{\tilde s}^2=
U\lambda(dx_1^2+dx_2^2)$. Here the order of operators in  a product is
important and a function represents the operation of multiplying on the left by
that function.
\begin{theorem}\label{quantumstackel} \begin{enumerate}\item $$[{\tilde
H},{\tilde K}]=0\iff
    [H,K]=0.$$
\item $${\tilde K}=\sum_{ij}\frac{1}{
    U\lambda}\partial_i\left((a^{ij}-\delta^{ij}\frac{W_U}{
      U\lambda}) U\lambda\right)\partial_j +\left(W-\frac{W_UV}{U}\right).$$
\end{enumerate}
\end{theorem}

\medskip\noindent PROOF:
\begin{enumerate}\item This is a straightforward verification, using
  the identities
\be\label{2ordident} [H_0,K_2]=0,\quad [H_0+V,K_2+K_0]=0,\quad [H_0+U,K_2+K_0^U]=0,\ee
$$ [A,BC]=B[A,C]+[A,B]C,\quad
[A,U^{-1}]=-U^{-1}[A,U]U^{-1}
$$
for linear operators $A,B,C$ and nonzero function $U$.
\item This follows from the fact that $\partial_i K_0^U\equiv \partial_i
W^U=\lambda \sum_j
  a^{ij}\partial_jU$.
\end{enumerate}
Q.E.D.

Note that the second part of the theorem shows that $\tilde K$ is indeed
formally self-adjoint on the manifold with metric $U\lambda (dx_1^2+dx_2^2)$.
Another way to see this is to use the formal definition of adjoint. With respect to the inner product on the space with weight function $U\lambda$ we have
$<f,g>_{U\lambda}=<f,Ug>_\lambda=<Uf,g>_\lambda$.
Thus
$$<{\tilde K}f,g>_{U\lambda}=<(UK-UK_0^U U^{-1}H)f,g>_\lambda=<f,(KU-HK_0^U)g>_\lambda$$
$$=<f,(U^{-1}KU-U^{-1}HK_0^U)g>_{U\lambda}.$$
This shows that  ${\tilde K}^*=U^{-1}KU-U^{-1}HK_0^U={\tilde K}$, where the final equality follows directly from identities (\ref{2ordident}).
\begin{corollary} If $K^{(1)},K^{(2)}$ are 2nd order symmetry
  operators for $H$, then
$$ [{\tilde K^{(1)}}, {\tilde K^{(2)}}]=0\iff [K^{(1)},K^{(2)}]=0.$$
\end{corollary}

Since one can always add a constant to a potential, it follows that $1/U$
defines an inverse St\"ackel transform of $\tilde H$ to $H$.

\subsection{3rd order operator symmetries}
A classical 3rd order symmetry takes the form  ${\cal K}={\cal K}_3+{\cal K}_1$
where
$${\cal K}_3=\sum ^2_{k,j,i=1}a^{kji}({\bf x})p_kp_jp_i,\quad  {\cal K}_1=\sum
^2_{\ell=1}b^\ell({\bf x})p_\ell.$$
The conditions $\{{\cal K}_3,{\cal H}_0\}=0$ are
\bea\label{3rdordersymalg1q}
2 a^{iii}_i &=&-3\left((\ln\lambda)_i a^{iii}+(\ln\lambda)_j a^{jii}\right),\
i\ne j\\
3 a^{jii}_i +
a^{iii}_j&=&-3\left((\ln\lambda)_ia^{iij}+(\ln\lambda)_j a^{ijj}\right)       
,\quad i\ne j\nonumber\\
2\left( a^{122}_1+a^{112}_2\right)&=&-(\ln\lambda)_1 a^{122}-(\ln\lambda)_1
a^{111}
-(\ln\lambda)_2 a^{222}-(\ln\lambda)_2 a^{112},\nonumber
\eea
which are just the requirements that the $a^{kji}$ be the components
of a 3rd order Killing tensor. The conditions $\{{\cal K}_3,V\}+\{{\cal
K}_1,{\cal H}_0\}=0$ are
\be
 b^1_2+ b^2_1   = 3
\sum_{s=1}^2\lambda a^{s21}
V_s,
\label{vectpot1}
\ee
$$ b^j_j=\frac32 \sum_{s=1}^2\lambda a^{sjj}
V_s-\frac12\sum_{s=1}^2(\ln\lambda)_s
b^s,
 \quad j=1,2, $$
and the condition $\{{\cal K}_1,V\}=0$ is
\be\label{scalpot11}\sum_{s=1}^2b^{s}
V_s=0.
\ee

Now let's consider a 3rd order operator symmetry $K$ that is skew-adjoint. The detailed conditions $[K,H]=0$ are complicated. However, we will restrict ourselves to systems with potentials that 
simultaneously admit a 3rd order classical symmetry and the related 3rd order quantum symmetry. A characteristic feature of such systems, and one that we will exploit, is that if $U$ is such a potential then so is $\alpha U$ for all scalars $\alpha$.  If $\cal K$ is the classical symmetry then we can write the operator symmetry in the form $K=K'_3+K'_1$ where  $K_3',K_1'$ are skew-adjoint of respective orders 3 and 1, 
$$K_3'=\sum
^2_{k,j,i=1}\left(a^{kji}\partial_{kji}+\frac{3}{2\lambda}(a^{kji}\lambda)_i\partial_{kj}
  +\frac{1}{2\lambda}(a^{kji}\lambda)_{kj}\partial_{i}\right),
$$
$$ K_1'=
\sum_{i=1}^2\left({ B}^i\partial_i+\frac{1}{2\lambda}({ B}^i\lambda)_i\right),
 \label{3rdorderconstofmot1q}
$$
and the terms $a^{kji}$ satisfy (\ref{3rdordersymalg1q}).

Now replace $V$ by $\alpha U$. Then the symmetry condition is $[K(\alpha),H_0+\alpha U]=0$ for all $\alpha$ where $K=K'_3+K'_1(\alpha)$. We assume that $K(\alpha)$ is analytic in $\alpha$ about $\alpha=0$. Then, $K'_3$ is independent of $\alpha$ and the dependence of $K'_1$ on $\alpha$ is at most 1st order. Thus we can write $B^i(\alpha)=c^i+\alpha b^i$ or $K'_1(\alpha)= K''_1+\alpha K_1$. The symmetry condition can be written as 
$$0=[K'_3+K''_1+\alpha K_1,H_0+\alpha U]$$
$$=[K'_3+K''_1,H_0]+\alpha([K'_3+K''_1,U]+[K_1,H_0])+\alpha^2 [K_1,U],$$
for all $\alpha$.
Setting $ K_3=K'_3+K''_1$ we have the identities
\be\label{3opident} [K_3,H_0]=0,\quad [K_3,U]+[K_1,H_0]=0,\quad [K_1,U]=0.\ee
Note that the 2nd order terms in the 2nd identity  are precisely the classical conditions (\ref{vectpot1}). The 3rd identity is precisely the classical condition (\ref{scalpot11}).
The operator $K_1''$ determines the transition from the classical constant of the motion to the operator symmetry. 
Now define
$${\tilde K}=K_3-K_1 U^{-1}(H_0+b),$$
where the operator order is important and $b$ is a constant.. A straightforward computation using identities, (\ref{3opident}) yields
$ [{\tilde K},U^{-1}(H_0+b)]=0$,
so  $\tilde K$ is a 3rd symmetry operator  for the Hamiltonian $U^{-1}(H_0+b)$.

\begin{theorem}\label{3rdorderthm} Let $H(\alpha)=H_0+\alpha U$, let $K(\alpha)$ be a 3rd order skew-adjoint symmetry of $H$, analytic in $\alpha$ about $\alpha=0$. Then there are 1st and 3rd order skew-adjoint operators $K_1,K_3$ such that $K(\alpha)=K_3+\alpha K_1$ and identities (\ref{3opident}) are satisfied. The operator ${\tilde K}=K_3-K_1^U U^{-1}(H_0+b)$ is a 3rd order symmetry for the system ${\tilde H}=U^{-1}(H_0+b)$.
\end{theorem}
\begin{corollary} ${\tilde K}^*=-{\tilde K}$ so $\tilde K$ is a 3rd order formally skew-adjoint symmetry of $\tilde H$.
\end{corollary}

\medskip \noindent  PROOF: This is a consequence of $ K^*=-K, {\tilde H}^*={\tilde H}$ and relations (\ref{3opident}). Q.E.D.

Note: The preceding argument  has to be modified in the special case that the system admits a 1st order $\alpha$-independent symmetry $L$: $[L,H(\alpha)]=0$. Then $K'_1(\alpha)$ need not be at most 1st order as a polynomial in $\alpha$. Indeed we can add a term $f(\alpha)L$ to $K'_1$ without changing the commutation relations. However, the conclusion  (\ref{3opident}) remains correct.

\begin{example} \label{91os} (The 9-1 anisotropic oscillator)  Let $H(\alpha)=\partial_{11}+\partial_{22}+\alpha(9x_1^2+x_2^2)$. This is a superintegrable system with generating 2nd and 3rd order symmetries
 $$ L=\partial_{22}+\alpha x_2^2,\quad K=\{x_1\partial_2-x_2\partial_1,\partial_{22}\} +\frac{\alpha}{3}(\{x_2^3,\partial_1\}-9\{x_1x_2^2,\partial_2\}),$$ where $\{S_1,S_2\}\equiv S_1S_2+S_2S_1$.
Let $U=(9x_1^2+x_2^2) +c$. It follows that the system 
$$ {\tilde H}= \frac{1}{(9x_1^2+x_2^2) +c}\left(\partial_{11}+\partial_{22}+b\right)$$
is superintegrable with one 2nd and one 3rd order symmetry.
\end{example}

\subsection{4th order operator symmetries}
Next we consider the case of a 4th order constant of the motion
\be
{\cal K}=\sum ^2_{\ell,k,j,i=1}a^{\ell kji}({\bf x})p_\ell p_kp_jp_i+\sum ^2_{m,q=1}b^{mq}({\bf x})p_mp_q+W({\bf x})={\cal K}_4+{\cal K}_2+{\cal K}_0,
 \label{4rthorderconstofmot1}
\ee
This  must satisfy   
the conditions 
\bea\label{4rthordersymalg1}
 a^{iiii}_i
  &=&-2\sum_{s=1}^2a^{siii}( \ln\lambda)_s\\
4 a^{jiii}_i+a^{iiii}_j&=&-6\sum_{s=1}^2a^{siij}(\ln\lambda)_s,\quad i\ne j\nonumber\\
3a^{jjii}_i+2a^{iiij}_j&=&-\sum_{s=1}^2a^{siii}(\ln\lambda)_s-3\sum_{s=1}^2a^{sijj}(\ln\lambda)_s,\quad i\ne j
\eea

\be
\label{tenspot2}
2 b^{ij}_i + b^{ii}_j 
 =6\lambda\sum_{s=1}^2a^{sjii}V_s-\sum_{s=1}^2b^{sj}(\ln\lambda)_s,\quad i\ne j
\ee
$$ b^{ii}_i  =2\lambda\sum_{s=1}^2a^{siii}V_s-\sum_{s=1}^2b^{sj}(\ln\lambda)_s,
$$
and
\be\label{scalpot3}\lambda\sum_{s=1}^2b^{si}V_s=W_i.
\ee
Note that the $a^{\ell kji}$ is  a 4th  order Killing tensor.

If $K$ is a 4th order symmetry operator, there exist
  functions $a^{\ell kji},{\tilde b}^{ij},{\tilde W}$ such that $K$ has the unique  self-adjoint form 
\be\label{4thorderconstofmot1q}
K=\sum
^2_{\ell,k,j,i=1}\frac{1}{\lambda}\partial_{ij}\left(a^{\ell kji}\lambda \partial_{k\ell}\right)
+\sum_{i,j=1}^2\frac{1}{\lambda}\partial_i\left({\tilde b}^{ij}\lambda \partial_j\right) +{\tilde W}=K'_4+K'_2+K'_0,
\ee
where the functions ${\tilde b}^{ij}(x_1,x_2),{\tilde W}(x_1,x_2)$ contain the parameter dependence.
Equating coefficients of  the  5th derivative terms in the
  operator condition 
$
[K,H]=0
$ 
  we obtain exactly the Killing tensor conditions (\ref{4rthordersymalg1}). 

 The remaining conditions on $K$ intertwine
  $\lambda,a^{\ell kji},{\tilde b}^{ji}$, ${\tilde W}$ and $V$, and are  complicated. Rather
  than solve them directly, we use the fact that the system with potential $\alpha U$ must be solvable for all  $\alpha$, and require that the symmetry $K(\alpha)$ is analytic in $\alpha$ about $\alpha=0$. The following argument for the form of $K$ is correct, up to addition of operators $f(\alpha) L_2$ or $g(\alpha)$ where $L_2$ is a 2nd order self-adjoint $\alpha$-independent  symmetry operator. Modulo this remark,  $K_1'(\alpha)$ must be at most a 1st order polynomial in $\alpha$ and $K'_0(\alpha)$ must be at most quadratic.  We can make the unique decomposition
 $${\tilde
    b}^{ji}( {\bf x})=c^{ji}({\bf x})+\alpha b^{ji}({\bf x})$$
$${\tilde W}= U^{(0)}({\bf x})+\alpha U^{(1)}({\bf x})+\alpha^2 W({\bf x}).$$
Substituting into $[K(\alpha),H_0+\alpha U]=0$ and equating the  3rd derivative terms that are linear in $\alpha$, we get exactly conditions (\ref{tenspot2})
Equating the coefficients of the 0th derivative terms that are quadratic in $\alpha$ we get exactly conditions (\ref{scalpot3}).

Now we write $K_2'(\alpha)=A_2+\alpha B_2, $, $K_0'(\alpha)=A_0+\alpha B_0 +\alpha^2 C_0$. It follows that
\be\label{4opident}  [K_4,H_0]=0,\quad [K_4,U]+[K_2,H_0]=0,\quad [K_2,U]+[K_0,H_0]=0,\ee 
where  
$$K=K_4+K_2+K_0,\quad K_4=K_4'+A_2+A_0,\quad K_2=B_2+B_0,\quad K_0=C_0.$$
Now define
$${\tilde K}=K_4-K_2 U^{-1}(H_0+b)+K_0\left(U^{-1}(H_0+b)\right)^2,$$
where the operator order is important. A straightforward computation using identities, (\ref{4opident}) yields
$ [{\tilde K},U^{-1}(H_0+b)]=0$.
Thus $\tilde K$ is    a 4th order symmetry operator for the Hamiltonian $U^{-1}(H_0+b)$.

\begin{theorem}\label{4thorderthm} Let $H(\alpha)=H_0+\alpha U$, let $K(\alpha)$ be a 4th order self-adjoint symmetry of $H(\alpha)$, analytic at $\alpha=0$.  Then there are 0th, 2nd and 4th order self-adjoint operators $K_0,K_2,K_4$ such that $K(\alpha)=K_4+\alpha K_2+\alpha^2 K_0$ and identities (\ref{4opident}) are satisfied. The operator ${\tilde K}=K_4-K_2 U^{-1}(H_0+b)+K_0( U^{-1}(H_0+b))^2$ is a 4th order symmetry for the system ${\tilde H}=U^{-1}(H_0+b)$.
\end{theorem}

\begin{corollary} ${\tilde K}^*={\tilde K}$ so $\tilde K$ is a 4th order formally self-adjoint symmetry of $\tilde H$.
\end{corollary}

\begin{example} \label{example5} This is an extension of Example \ref{example3} to the quantum case, \cite{TTW}.
Let
$$ { H}=\partial_{11}+\partial_{22}+a(x_1^2+x_2^2)+b \frac{(x_1^2+x_2^2)}{ (x_1^2-x_2^2)^2} +c 
\frac{(x_1^2+x_2^2)}{ x_1^2x_2^2}$$
There are  two basic self-adjoint symmetry operators, one of 2nd order,
$${ K}_2=(x_1\partial_2-x_2\partial_1)^2+4b \frac{x_1^2x_2^2}{(x_1^2-x-2^2)^2} + c \frac{(x_1^4+x_2^4)}{ x_1^2x_2^2}$$
and one of 4th order,
$${ K}_4=(\partial_{11}-\partial_{22})^2+[2ax_1^2+ 2b \frac{(x_1^2+x_2^2)}{(x_1^2-x_2^2)^2} -2c 
\frac{(x_1^2-x_2^2)}{ x_1^2x_2^2} ]\partial_{11}$$
$$+[-4ax_1x_2+ 8b \frac{x_1x_2}{ (x_1^2-x_2^2)^2} ]\partial_{12}+[2ax_2^2+ 2b 
\frac{(x_1^2+x_2^2)}{ (x_1^2-x_2^2)^2} +2c \frac {(x_1^2-x_2^2)}{ x_1^2x_2^2} ]\partial_{22}$$
$$+(2ax_1-\frac{4c}{x_1^3})\partial_1+(2ax_2-\frac{4c}{x_2^3})\partial_2 +a^2(x_1^2-x_2^2)^2+ \frac{b^2}{ (x_1^2-x_2^2)^2}$$
$$ + c^2 \frac{(x_1^2-x_2^2)^2}{ x_1^4x_2^4}+8ab 
\frac{x_1^2x_2^2}{ (x_1^2-x_2^2)} +2 \frac{bc}{ x_1^2x_2^2}+6c(\frac{1}{x_1^4}+\frac{1}{x_2^4}).$$
These operators generate a closed symmetry algebra. Let
 ${ R}=[{ K}_2,{ K}_4]$.The  relations are 
$$[{ K}_2,{ R}]=32{ H}^2K_2-32\{ K_4, K_2\}-64(b+2c+4){K}_4+64(b-c+2){ H}^2$$
$$-128a(b+1){ K}_1-128a(b^2+2bc+4b+6c+4),$$
$$[ K_4,{ R}]=32 K^2_4-32 H^2K_4+128a{ K}_2{ H}^2-384a^2{ K}^2_2+128a(b+1){ K}_4-64a(b+4c+6){ H}^2$$
$$+256a^2
(2c-b+14){ K}_2 + 128a^2(b^2+4c^2+20bc+18b+8c-8)$$
There is also a Casimir constraint. Then the St\"ackel transformed system
$${\tilde { H}}=\frac{1}{(x_1^2+x_2^2)+B \frac{(x_1^2+x_2^2)}{ (x_1^2-x_2^2)^2} +C 
\frac{(x_1^2+x_2^2)}{ x_1^2y_2^2}+D } \left(\partial_{11}+\partial_{22}+\right.$$
$$\left. a(x_1^2+x_2^2)+b \frac{(x_1^2+x_2^2)}{ (x_1^2-x_2^2)^2} +c 
\frac{(x_1^2+x_2^2)}{ x_1^2x_2^2}+d\right)$$
is also superintegrable with 4th and 2nd order generating self-adjoint symmetries.
 \end{example}

\subsection{$N$th order operator symmetries}
A possible structure of the $N$th order operator case is now clear, though it is far from clear whether the structure includes all cases. Suppose the system  $H(\alpha)=H_0+V+\alpha U\equiv H+\alpha U$ admits a truly $N$th order symmetry operator $K(\alpha)$ analytic in $\alpha$ about $\alpha=0$, where $N\ge 2$ and   $K$  is self-adjoint for even $N$, skew-adjoint for odd $N$. Then we can write
$$K(\alpha)=K'_N +\sum_{j=1}^{[n/2]}K'_{N-2j}(\alpha)$$
where each $K'_{N-2j}$ is self-adjoint or skew-adjoint, depending on the parity of $N$. The symmetry condition is
\be\label{Nthordsymmcond1} [K'_N+\sum_{j=1}^{[N/2]} K'_{N-2j}(\alpha),H+\alpha U]=0,\ee
where the $K'_{N-2j}(\alpha)$ are analytic in $\alpha$. Suppose, modulo terms of the form $f^{(i)}(\alpha)L_{N-2j}$ where $L_{n-2j}$ is an $\alpha$-independent symmetry of $H+\alpha U$ for $j>0$, we  have
$$  K'_{N-2j}=\sum_{i=0}^{j}A_{N-2j}^{(i)}\alpha^i,\quad j=0,1,\cdots, [N/2]$$
where the $A_{N-2j}^{(i)}$ are independent of $\alpha$. Setting $K_{N-2j}= \sum_{h=0}^{[(N-2j)/2]} A^{(j)}_{N-2j-2h}$ we have $K=\sum_{j=0}^{[N/2]}K_{N-2j}$ and the symmetry condition (\ref{Nthordsymmcond1}) becomes
\be\label{Nthordsymmcond2} [\sum_{j=0}^{[N/2]} \alpha^j K_{N-2j},H+\alpha U]=0,\ee
or 
\be \label{Nthordsymmcond3}[ K_{N-2j},U]+[ K_{N-2j-2}, H]=0,
\quad j=0,1,\cdots,[N/2],\ee
where we define $K_{N-2j}\equiv 0$ for $j>[N/2]$ and $j<0$.

Now define
$${\tilde K}=\sum_{h=0}^{[N/2]}(-1)^h K_{N-2j} \left(U^{-1}(H+b)\right)^h,$$
where $b$ is a constant. 
From relations (\ref{Nthordsymmcond3}) we have
\be\label{Nident}[{\tilde K},U^{-1}(H+b)]=\sum_{h=0}^{[N/2]}(-1)^h[K_{N-2h},U^{-1}(H+b)]\left(U^{-1}(H+b)\right)^h=\ee
$$U^{-1}\sum_{h=0}^{[N/2]}\left([K_{N-2h},U](-1)^{h+1}(U^{-1}(H+b))^{h+1}+[K_{N-2h},H](-1)^h(U^{-1}(H+b))^h\right)$$
$$= U^{-1}\sum_{h=1}^{[N/2]}(-1)^h\left([K_{N-2h+2},U]+[K_{N-2h}^U,H]\right)(U^{-1}(H+b))^h=0.$$
 Thus $[{\tilde K},{\tilde H}]=0$.

\begin{theorem}\label{Nthorderthm} Let $H(\alpha)=H_0+V+\alpha U \equiv H+\alpha U$ and $N\ge 2$. Let $K(\alpha)$ be a nonzero $N$th order operator  symmetry of $H(\alpha)$ analytic at $\alpha=0$,  self-adjoint for even $N$ and skew-adjoint for odd $N$. Suppose further there are  operators $K_{N-2j}$  such that $K(\alpha)=\sum_{j=0}^{[N/2]} K_{N-2j}\alpha^j$ and identities (\ref{Nthordsymmcond2}), (\ref{Nthordsymmcond3}) are satisfied.  Then the operator ${\tilde K} =\sum_{h=0}^{[N/2]}(-1)^h K_{N-2h} (U^{-1}(H+b))^h$ is an $N$th order symmetry for the system ${\tilde H}=U^{-1}(H+b)$.
\end{theorem}

\begin{example} \label{91osgen} (The 9-1 anisotropic oscillator) This is a generalization of Example \ref{91os} to a full St\"ackel transform.  Let $H(0)=\partial_{11}+\partial_{22}+a(9x_1^2+x_2^2)$  and $L$ be as in Example  \ref{91os} with $\alpha$ replaced by $a+\alpha$, and  $U=(9x_1^2+x_2^2) +c$. It follows that the system 
$$ {\tilde H}= \frac{1}{(9x_1^2+x_2^2) +c}\left(\partial_{11}+\partial_{22}+a(9x^2+y^2)+ b)\right)$$
is superintegrable with one 2nd and one 3rd order symmetry.
\end{example}

Note that Theorem \ref{Nthorderthm} does not require that the quantum system go to a classical system, only that a scalable potential term can be split off. Thus it applies to ``hybrid'' quantum systems that have a classical part.
\begin{example} \label{91qos} (The hybrid  9-1 anisotropic oscillator)  Let $H(0)=\partial_{11}+\partial_{22}+a(9x_1^2+x_2^2)-2/x_2^2$. This is a superintegrable system with generating 2nd and 3rd order symmetries,
 $$ L=\partial_{22}+ax_2^2,\quad K=\{x_1\partial_2-x_2\partial_1,\partial_{22}\} +\{\frac{a}{3}x_2^3+\frac{1}{x_2},\partial_1\}-\{3x_1(ax_2^2+\frac{1}{x_2^2}),\partial_2\}.$$  Note that this system does not have a classical limit. [Using a different normalization that makes clear the classical limit, Gravel writes this Hamiltonian as $H(0)=-(\hbar^2/2)(\partial_{11}+\partial_{22})+a(9x_1^2+x_2^2)+{\hbar^2}/x_2^2$.]
 Let\  $U=(9x_1^2+x_2^2) +c$. It follows that the system 
$$ {\tilde H}= \frac{1}{(9x_1^2+x_2^2) +c}\left(\partial_{11}+\partial_{22}+a(9x_1^2+x_2^2)-\frac{2}{x_2^2}+b\right)$$
is superintegrable with one 2nd and one 3rd order symmetry.
\end{example}

\begin{example} \label{91qost} (A translated hybrid  9-1 anisotropic oscillator)  This is a slight modification of Example \ref{91qos}. Let $H(0)=\partial_{11}+\partial_{22}+a(9x_1^2+x_2^2)+cx_1-2/x_2^2$. This is a superintegrable system with generating 2nd and 3rd order symmetries, and no classical limit.
 Let\  $U=x_1$. It follows that the system 
$$ {\tilde H}= \frac{1}{x_1}\left(\partial_{11}+\partial_{22}+a(9x_1^2+x_2^2)+cx_1-\frac{2}{x_2^2}+b\right)$$
is superintegrable with one 2nd and one 3rd order symmetry. This space is a Darboux space of type 1, \cite{KKMW2003}.
\end{example}

\begin{example}  Let $H(0)=\partial_{11}+\partial_{22}+a/x_1^2-2/x_2^2$. This is a superintegrable system with two linearly independent  2nd and three linearly independent 3rd order symmetries, \cite{Gravel}.
  This system does not have a classical limit. [Using a different normalization that makes clear the classical limit, Gravel writes this Hamiltonian as $H(0)=-(\hbar^2/2)(\partial_{11}+\partial_{22})+a/x_1^2+{\hbar^2}/x_2^2$.]
 Let\  $U=1/x_1^2 +c$. It follows that the system 
$$ {\tilde H}= \frac{x_1^2}{1 +cx_1^2}\left(\partial_{11}+\partial_{22}+\frac{a}{x_1^2}-\frac{2}{x_2^2}+b\right)$$
is superintegrable with with two linearly independent  2nd and three linearly independent 3rd order symmetries. In the case $c=0$ this is a superintegrable system on a space of nonzero constant curvature. Indeed, for $x_1,x_2$ real, it is the upper half space metric of non-Euclidean geometry.
 \end{example}

In the operator case where $V=0$ in Theorem \ref{Nthorderthm} there is always a corresponding classical system. 
Indeed, equations (\ref{Nthordsymmcond2}) and (\ref{Nthordsymmcond3}) clarify the close relationship between symmetries of quantum systems with potentials invariant under scaling and classical constants of the motion. If we set $\alpha=1/\hbar^2$, $V=0$ in (\ref{Nthordsymmcond2}) we can rewrite this expression as 
$$ [\sum_{j=0}^{[N/2]} \hbar^{N-2j}K_{N-2j},\hbar^2 H_0+U]=0.$$
Further if we write the differential terms in the operators $K_{N-2j}$ as
$$ K_{N-2j}=\sum_{i_s =1,2} a^{i_1\cdots i_{N-2j}}\partial_{i_1}\cdots\partial_{i_{N-2j}}+\ {\rm lower\ order\ terms},$$
we can associate these operators with the phase space functions 
$${\cal K}_{N-2j}({\bf x},{\bf p})=\sum_{i_s=1,2} a^{i_1\cdots i_{N-2j}} p_{i_1}\cdots p_{i_{N-2j}}.$$
Then by equating coefficients of the highest order derivative terms in equations (\ref{Nthordsymmcond3}) we obtain
the Poisson bracket relations 
\be \label{Nthordfcncond3}\{ {\cal K}_{N-2j},U\}+\{ {\cal K}_{N-2j-2}, {\cal H}_0\}=0,
\quad j=0,1,\cdots,[N/2],\ee
so that ${\cal K}=\sum_{j=0}^{[N/2]} {\cal K}_{N-2j}$ is an $N$th order constant of the motion for the system with Hamiltonian ${\cal H}={\cal H}_0+U$.

\section{Conclusions and outlook}
We have found specializations of classical CCM that preserve the order of symmetries and determine symmetry algebra homomorphisms, and for 2D manifolds we have extended them to the quantum case. Generally speaking, these transforms apply to systems with a nonconstant potential that admits scaling in at least one parameter. They do not apply to quantum systems with no classical counterpart in which the potential is fixed.  This tool makes it clear that superintegrable systems occur for a wide variety of manifolds, not just on constant curvature spaces.  For 2nd order superintegrable systems the St\"ackel transform has been used effectively in 2D  to show that all such systems are transforms of systems on constant curvature spaces, and this has lead to an elegant classification of all such systems. It is our aim to develop CCM to investigate the possibility of a similar classification for 3rd and higher order superintegrable systems.

For simplicity, we have restricted our quantum constructions  to 2D manifolds though some partial results hold in $n$ dimensions. There appears to be no insurmountable barrier to extending these results to 3D and higher conformally flat manifolds, but the details have not yet been worked out. Clearly gauge transformations are required and the gauge will be a function of the scalar curvature of the manifold.

\end{document}